\documentclass[amsmath,amssymb,aps,prd,twocolumn,superscriptaddress,showpacs]{revtex4}
\usepackage{graphicx}
\usepackage{amsmath,amsthm,amssymb}
\usepackage{cases}
\usepackage{times}
\usepackage{color}

\newcommand{\Comments}[1]{}
\newcommand{\PB}[1]{\left\lbrace{#1}\right\rbrace_\mathrm{PB}}
\newcommand{\KB}{Kadanoff-Baym}
\begin{document}
\title{Entropy current for the relativistic Kadanoff-Baym equation and H-theorem
in $O(N)$ theory with NLO self-energy of $1/N$ expansion}
\author{A. Nishiyama}
\affiliation{Yukawa Institute for Theoretical Physics,
Kyoto University, Kitashirakawa-Oiwakecho, Sakyo, Kyoto 606-8502, Japan}
\author{A. Ohnishi}
\affiliation{Yukawa Institute for Theoretical Physics,
Kyoto University, Kitashirakawa-Oiwakecho, Sakyo, Kyoto 606-8502, Japan}

\begin{abstract}
We derive an expression of
the kinetic entropy current in the nonequilibrium $O(N)$ scalar theory
from the Schwinger-Dyson (Kadanoff-Baym) equation
with the 1st order gradient expansion. 
We show that our kinetic entropy satisfies the H-theorem
for the leading order of the gradient expansion
with the next-to-leading order self-energy of the $1/N$ expansion
in the symmetric phase, 
and that entropy production occurs
as the Green's function evolves with an nonzero collision term. 
Entropy production stops at local thermal equilibrium
where the collision term contribution vanishes
and the maximal entropy state is realized.
Next we also compare our entropy density
with that in thermal equilibrium which is given from thermodynamic potential 
or equivalently 2 particle irreducible effective action.
We find that our entropy density corresponds
to that in thermal equilibrium with the next-to-leading order skeletons of
the $1/N$ expansion if skeletons with energy denominators in momentum integral
can be regularized appropriately.
We have a possibility that memory correction terms remain
in entropy current if not regularized. 
\end{abstract}
\maketitle

\section{Introduction}

The 2 Particle Irreducible (2PI) effective functional technique
provides a powerful tool to 
deal with controlled nonequilibrium dynamics with non-secularity 
and late time universality \cite{BergesReview,BergesSerreau}.
It gives the Kadanoff-Baym (KB) equation which describes the dynamics for
quantum fluctuations of fields.
In 1960s, 
Baym and Kadanoff studied the Schwinger-Dyson (SD) equation
for the two-point function $G(x,y)$ for nonequilibrium systems~\cite{BK,KB62}
on the basis of a functional approach developed by Luttinger and Ward \cite{LW}.
Then Baym reformulated it in terms of variational principle,
introducing the so-called $\Phi$-derivable approximation~\cite{Baym}
which is given by a truncated set of closed 2PI diagrams. 
The main virtue of this approximation is that 
the resulting equations conserve the charge, energy and momentum of the system.
This approach has been reformulated
by use of path integral method~\cite{Schwinger,Keldysh,CJT,CH}. 
It is applied to a variety of areas in physics,
such as cosmology, ultrarelativistic heavy ion collisions,
or condensed matter physics.
In describing the reheating processes during the early stage of inflation,
it becomes necessary to trace the time 
evolution of slowly evolving inflaton field and quantum
fluctuations~\cite{BS2003,AT2008,Tranberg2008,AST2004}.
In condensed matter physics it can be applied to Bose-Einstein Condensate (BEC),
since 2PI approach is a candidate with properties of gapless excitation
and conservation laws~\cite{vHK2002-3}. 
The 2PI approach with the KB equation would be useful also in understanding
the early thermalization processes towards quark-gluon plasma formation
in high-energy heavy-ion collisions~\cite{BBW2004},
while it may be necessary to combine the classical field dynamics
leading to the linear rise of entropy from the chaotic nature of the
system~\cite{KMOS2009}.


In this paper we focus on the 2PI approach to the scalar $O(N)$ theory,
where identical bosons interact one another in the symmetric phase.
The $O(N)$ model has been employed
in the vacuum and thermal equilibrium analyses
and it is also applied to time dependent phenomena~\cite{BergesReview}.
It can be used in inflationary processes of the early universe,
the formation of Bose-Einstein condensate in the laboratory,
and the chiral phase transition in heavy ion collisions.
The dynamics of the chiral phase transition following the expansion
of a quark-gluon plasma produced in relativistic heavy ion collisions
has been analyzed by an $O(4)$ $\sigma$ model
in the leading order of $1/N$ expansion~\cite{CKMP,LDC}.

In the analysis of the $O(N)$ model,
the large $N$ approximation (or the $1/N$ expansion) has been applied
since 1960s or 1970s
in both statistical mechanics
and quantum field theory~\cite{Stanley,Wilson,CJP}.
The $1/N$ expansion has the advantage over the loop expansion
that it is not restricted to small couplings.
However a naive diagrammatic 
$1/N$ expansion breaks down
when it is applied to the dynamical simulations,
because a secular (unbounded) time evolution prevents the description
of the late-time behavior of quantum field.
As an improved treatment,
we can take account of the self-consistency
in the 1 particle irreducible (1PI) effective action,
which corresponds to the first Legendre transformation
of the generating functional with respect to the expectation value of the field,
but the 1PI effective action is also plagued
by the secular problems and unitarity violation;
for example 
$\langle \phi(x)^2 \rangle$ can become negative
in the late time behavior~\cite{MDC,MACDH}.
Therefore in order to extract the stable evolution
without the above secularity we need to go beyond the 1PI technique,
so that we adopt 2PI technique as the simplest example.
Recently a systematic $1/N$ expansion of the 2PI effective action
has been applied to a scalar $O(N)$ in the symmetric
and broken phases~\cite{AABBS}.
By resuming all 2 particle reducible diagrams with higher order secularity
with respect to time,
2PI approach with the KB equation realizes the controlled time evolution
without the secular problem and the unitarity violation
in far-from-equilibrium dynamics~\cite{Berges}. 

Starting from controlled numerical analyses of the next-to-leading order (NLO)
skeletons of $1/N$ expansion of the $O(N)$ model in $1+1$ dimensions
with vanishing classical field in Ref.~\cite{Berges},
simulations in the broken phase in Refs.~\cite{AT2008,AST2004}
and estimation of nonthermal fixed point~\cite{BRS} have been also done.
It is numerically found that we need to include the NLO contribution of the
$1/N$ expansion in order to describe quantum scattering and thermalization.
Recently NNLO of the $1/N$ expansion is derived~\cite{AartsTranberg2006}
and the rapid convergence property for moderate values of $1/N$
is observed~\cite{ALT2008}.
Numerical analyses show 
the thermalization of the distribution function
derived from the two-point Green's function for the truncated self-energy.
However there is no consideration of kinetic entropy
and its H-theorem based on the KB equation.
Thus we concentrate to derive the
analytic expression of the entropy current $s^\mu$
in nonequilibrium from KB equation
and to show that the H-theorem is satisfied at the level of the Green's function
analytically for the NLO self-energy of the $1/N$ expansion 
in the symmetric phase.
In the end, we aim to give a criteria whether thermalization occurs or not
in off-shell propagation for given self-energy or collision processes. 
In $\phi^4$ theory with the NLO self-energy of the coupling expansion, 
numerical simulations without gradient expansion
in Refs.~\cite{AB,BC,JCG,AST,LM} shows the resultant thermalization. 
The proof of the H-theorem in Refs.~\cite{IKV4,Kita,Nishiyama}
is consistent with these numerical results. 
The work given in this paper is an extension of our recent work
on the H-theorem in $\phi^4$ theory to $O(N)$ theory~\cite{Nishiyama}.
We show the H-theorem for local parts of NLO self-energy of the $1/N$ expansion
in the $O(N)$ theory.
We find that the resultant H-theorem is consistent with numerical analyses
performed for the $O(N)$ theory.
Judging from this consistency, the proof of the H-theorem might be a criterion
to confirm whether thermalization occurs or not without numerical simulation.
It might be possible to investigate qualitative properties of entropy production
in quantum field theories such as non-Abelian gauge theory,
which we still have difficulty in numerical estimate,
for example, due to singularities in massless quantum fluctuations.
 
We also give an analytic expression of the entropy density $\cal S$ 
in thermal equilibrium derived from the thermodynamic potential
(2PI effective action) with NLO skeletons of the $1/N$ expansion.
The present entropy density $s^0$ based on the H-theorem
should correspond to the entropy density $\cal S$ in thermal equilibrium.
We find that entropy density $s^0$ is consistent with the entropy density
$\cal S$ in thermal equilibrium
if we can regularize skeletons in higher loop order. 
If singularities remain, 
memory correction terms as shown in Ref.~\cite{IKV4} might appear.
Hence we have to estimate the singularities of skeltons carefully
order by order in practical applications.
 
In this paper first we review the Kadanoff-Baym equation
with the NLO self-energy of the $1/N$ expansion.
Next we derive an analytic expresson of the entropy current
for the relativistic field theory in the symmetric phase,
and show the H-theorem for the $O(N)$ theory in NLO of the $1/N$ expansion. 
In the derivation we adopt the gradient expansion with respect
to coordinate space, which is necessary to violate time reversal invariance
of the KB equation.
When we ignore higher order derivative terms,
spatial fluctuations in a small volume are assumed to vanish.
Thus the gradient expansion corresponds to a coarse graining procedure.
In the end we compare our entropy density $s^0$ for nonequilibrium systems
with entropy density $\cal S$ in thermal equilibrium. 

\section{Kadanoff-Baym equation with NLO self-energy of $1/N$ expansion}

\begin{figure}[tb]
\begin{center}
\includegraphics[width=6cm]{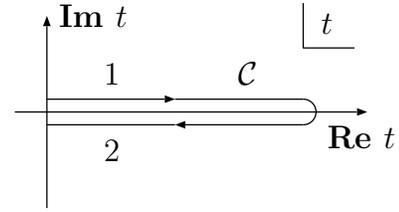}
\caption{
Integral path in the time coordinate.
}
\label{fig:path}
\end{center}
\end{figure}

In this section, we briefly review the Kadanoff-Baym equation 
with the NLO self-energy of the $1/N$ expansion in the $O(N)$ model.
We follow the notation in Ref.~\cite{BergesSerreau}. 
Let us consider a relativistic real scalar field
$\phi_a$ ($a=1,\cdot \cdot \cdot N$) with the $O(N)$ 
symmetric action,
\begin{eqnarray}
S[ \phi] = \int_x \left[\frac{1}{2} (\partial \phi_a)^2 -\frac{1}{2}m^2 \phi_a^2 -\frac{\lambda}{4! N} 
\left(  \phi_a \phi_a \right)^2 \right]
\ ,
\end{eqnarray}
where summation over indices is implied.
We adopt a closed time path $\cal C$ along the real time axis,
the path from $t_0$ to $\infty$ and from $\infty$ to $t_0$
as shown in Fig.~\ref{fig:path},
in order to trace nonequilibrium dynamics~\cite{Schwinger, Keldysh}. 
The 2PI effective action $\Gamma$ with vanishing mean field
$\langle \phi \rangle=0$ (symmetric phase) is written as
\begin{eqnarray}
\Gamma[G] &=&
\frac{i}{2} {\rm Tr} {\rm ln} \left( G \right) ^{-1} 
+ \frac{i}{2} {\rm Tr}G_0^{-1} G +\frac{1}{2} \Phi[ G]\, 
\label{Action}
\end{eqnarray}
with the full two-point Green's function 
\begin{align}
&G_{ab}(x,y) \equiv 
\langle {\rm T}_{\cal C}  \phi_a(x) \phi_b(y)\rangle
\nonumber\\
&= 
\theta_{\cal C}(x^0-y^0) \langle  \phi_a (x)  \phi_b (y) 
\rangle + \theta_{\cal C}(y^0-x^0) \langle  \phi_b (y)  \phi_a (x) \rangle
\nonumber\\
&= \delta_{ab}G(x,y). 
\end{align}
Here $G_{ab}(x,y)$ defined on the closed time path $\mathcal{C}$
can be also written in a matrix notation
\begin{eqnarray}
G_{ab}(x,y)=\left(
\begin{array}{cc}
 G^{11}_{ab}(x,y) & G^{12}_{ab}(x,y)   \\
 G^{21}_{ab}(x,y) & G^{22}_{ab}(x,y)   
\end{array}
\right) 
\ ,
\end{eqnarray} 
with 
\begin{subequations}
\begin{align}
G^{21}_{ab}(x,y) =& \langle  \phi_a(x)   \phi_b(y) \rangle,  \\
G^{12}_{ab}(x,y) =& \langle  \phi_b(y)  \phi_a(x) \rangle,  \\
G^{11}_{ab}(x,y)
  =&\theta(x^0-y^0) G^{21}_{ab}(x,y) + \theta (y^0-x^0) G^{12}_{ab}(x,y),  \\
G^{22}_{ab}(x,y)
  =& \theta(y^0-x^0) G^{21}_{ab}(x,y) + \theta (x^0-y^0) G^{12}_{ab}(x,y).
\end{align}
\end{subequations}
The upper label "1" represents the path from $t_0$ to $\infty$
and "2" represents that from $\infty$ to $t_0$ 
in the closed time path contour $\cal C$.
The term $\Phi[G]/2$ contains infinite series of 2PI diagrams
whose lines are given by the full Green's function $G$.
The stationary condition for the effective action (\ref{Action})
\begin{eqnarray}
\frac{\delta \Gamma}{\delta G}=0 
\ ,
\label{sc}
\end{eqnarray}
gives rise to the SD equation for the non-equilibrium
Green's function $G(x,y)$, {\em i.e.} the KB equation,
\begin{eqnarray}
G^{-1}(x,y) = G^{-1}_{0}(x,y) - \Sigma(x,y)
\ .
\label{KB}
\end{eqnarray}
The proper self-energy ($\Sigma$)
and the free Green's function ($G_{0}$) are defined as
\begin{align}
\Sigma=& i{\delta \Phi[G] }/{\delta G}
\ ,\\
iG_{0}^{-1}(x,y)=&-(\partial^2_x+m^2)\delta_{\cal C}(x-y)
\ ,
\end{align}
where $\delta_{ab}$ in front of $G$ and $\Sigma$ can be factorized.

\begin{figure}[bt]
\begin{center}
\includegraphics[width=8cm]{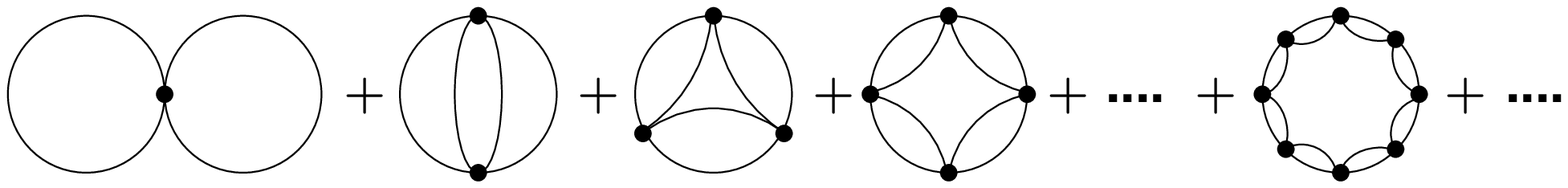}
\caption{NLO skeleton diagrams of $1/N$ expansion in symmetric phase~\cite{AABBS}. }
\label{fig:NLO}
\end{center}
\end{figure}

The NLO diagrams of $1/N$ expansion in $\Phi[G]/2$ is given by the series of diagrams 
shown in Fig.\ref{fig:NLO}, which can be written
in equations as~\cite{AABBS,Berges},
\begin{align}
&\frac{1}{2} \Phi[G]
=-\frac{\lambda }{4!N} \int_x G_{aa}(x,x) G_{bb}(x,x)
+ \frac{i}{2} {\rm Tr} \ln B[G]
\ ,\\
&B(x,y;G) \equiv \delta_{\cal C} ^{(d+1)} (x-y)
+ \frac{i\lambda}{6N}\, G_{ab}(x,y) G_{ab}(x,y) \ .
\end{align} 

The two-point correlation function $G(x,y)$ in the Schwinger-Keldysh formalism
is expressed by two independent functions;
the spectral function $\rho(x,y)$
and the statistical function $F(x,y)$,
\begin{align}
G^{12}(x,y)=&F(x,y)+\frac{i}{2}\rho(x,y)
\label{eq:FrhoA}
\ ,\\
G^{21}(x,y)=&F(x,y)-\frac{i}{2}\rho(x,y)
\label{eq:FrhoB}
\ .
\end{align}
Notice that $F(x,y)= F(y,x)$ and $\rho(x,y)=-\rho(y,x)$. 
The spectral function $\rho(x,y)$ contains the information
about which states are the most realized,
while the statistical function $F(x,y)$ determines
how much a state is occupied.
Then KB equation (\ref{KB}) can be reexpressed
by using $F(x,y)$ and $\rho(x,y)$ as
\begin{align}
&\left[ \partial _x + m^2 + \lambda \frac{N+2}{6N} F(x,x) \right] F(x,y)
\nonumber\\
=& -\int_{t_0}^{x^0}dz^0 d^{d}z 
\left[
 \hat \Sigma_\rho (x,z) F(z,y) - \hat \Sigma_F(x,z) \rho(z,y)
\right]
\label{eq:KBF}
\ ,\\
&\left[ \partial _x + m^2 + \lambda \frac{N+2}{6N} F(x,x) \right] \rho(x,y)
\nonumber\\
=& -\int_{y^0}^{x^0}dz^0 d^{d}z 
\hat \Sigma_\rho (x,z) \rho(z,y) 
\ .
\label{eq:KBR}
\end{align} 
Here the self-energies are also decomposed into 
the statistical and spectral parts,
\begin{align}
\Sigma^{12}(x,y)=&\hat \Sigma_F(x,y)+\frac{i}{2}\Sigma_\rho(x,y)
\label{eq:FrhoSA}
\ ,\\
\Sigma^{21}(x,y)=&\hat \Sigma_F(x,y)-\frac{i}{2}\Sigma_\rho(x,y)
\label{eq:FrhoSB}
\ .
\end{align}
These are given by 
\begin{align}
\hat \Sigma_F(x,y)
=& F(x,y) \hat D_F(x,y) -\frac{1}{4} \rho(x,y) \hat D_\rho(x,y)
\label{sigfab}
\ ,\\
\hat \Sigma_\rho(x,y)
=& \rho(x,y) \hat D_F(x,y) + F(x,y) \hat D_\rho(x,y) 
\ .
\label{sigrab}
\end{align} 
The functions $\hat D_F$ and $\hat D_\rho$ are given by resummation: 
\begin{align}
&\hat D_F(x,y) = \frac{\lambda}{3N} \left[-\hat \Pi_F(x,y) \right.
\nonumber\\
&\left.
- \int_{t_0}^{y^0} dz^0 d^d z
\left(
D_F (x,z) \hat \Pi_\rho (z,y) - D_\rho (x,z) \hat \Pi_F(z,y)
\right)\right],
\label{hdfxy}
\\
&\hat D_\rho(x,y) = \frac{\lambda}{3N} \left[-\hat \Pi_\rho(x,y)
+ \int_{t_0}^{y^0} dz^0 d^d z
\hat \Pi_\rho(x,z)\hat 
 D_\rho(z,y)\right],
\label{hdrxy}
\end{align} 
with the functions $\hat \Pi_F$ and $\hat \Pi_\rho$ given as,
\begin{align}
\hat \Pi_F(x,y)
=& \frac{\lambda}{6} \left[ F^2(x,y) -\frac{1}{4} \rho^2(x,y) \right]
\label{pifxy}
\,\\
\hat \Pi_\rho(x,y)
=& \frac{\lambda}{3} F(x,y) \rho(x,y)
\ .
\label{pirxy}
\end{align} 
Here notice that $\hat \Pi_F$ and $\hat \Pi_\rho$ represent
the chain part of the diagrams in Fig.~\ref{fig:NLO}.
In the limit of $N\rightarrow \infty$
the R.H.S. of Eqs.~(\ref{eq:KBF}) and (\ref{eq:KBR}) vanish
and neither thermalization nor damping of unequal-time Green's functions occur.

\section{H-theorem for $O(N)$ scalar field theory with NLO self-energy}
\subsection{Introduction of kinetic entropy}
\begin{figure}[tb]
\begin{center}
\includegraphics[width=8cm]{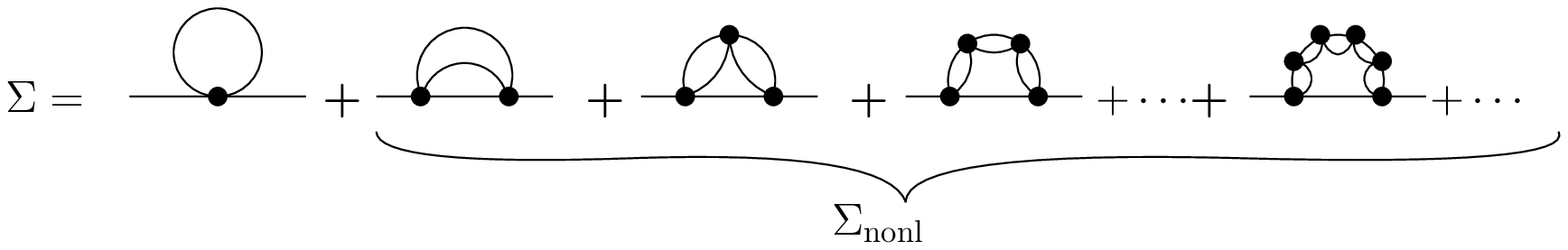}
\caption{Local and non-local part of the self-energy.}
\label{fig:nonl}
\end{center}
\end{figure}

In the derivation of kinetic entropy from the Kadanoff-Baym
(or equivalently Schwinger-Dyson) equation (\ref{KB}),
we adopt the 
$G^{12}$ and $G^{21}$ as independent functions in the Green's function.
We start from the \KB\ equation (\ref{KB}).
Multiplying $G$ from the right and left hand sides of Eq.~(\ref{KB}),
respectively, and by using $\int_z G^{-1}(x,z) G(z,y)=\delta(x-y)$,
we obtain,
\begin{align}
&-\left[
\partial_x^2+m^2+ \frac{\lambda(N+2)}{6N}G^{aa}(x,x) 
\right] G^{ab}(x,y)
\nonumber\\
&-i \int dz \Sigma_{\rm nonl}^{ac}(x,z)c^{cd}G^{db}(z,y) 
=
ic^{ab}\delta(x-y)
\ ,
\label{SDR}
\\
&-\left[
\partial_y^2+m^2+ \frac{\lambda(N+2)}{6N}G^{bb}(y,y)
\right]G^{ab}(x,y)
\nonumber\\
&-i\int dz 
G^{ac}(x,z)c^{cd} \Sigma_{\rm nonl}^{db}(z,y)
=
ic^{ab}\delta(x-y)
\; .
\label{SDL}
\end{align}
The integral over $z^0$ is from $t_0$ to $\infty$,
and the sign factor $c^{ab}={\rm diag}(1,-1)$ reflects the direction
of integral on branches 1 and 2.
The local part of the self-energy, shown in the first graph
in Fig.~\ref{fig:nonl}, is included in the first term
as the modification of the mass, and the non-local part
$\Sigma_{\rm nonl}$ is defined as the rest,
\begin{align}
\Sigma^{ab}(x,y)=&-i \frac{\lambda(N+2)}{6N}G^{ab}(x,y)\delta(x-y)\delta_{ab}
\nonumber\\
	&+ \Sigma^{ab}_{\rm nonl}(x,y)
\ .
\end{align}

In deriving the H-theorem,
we invoke the gradient expansion
in the the ``center-of-mass'' coordinate $X=(x+y)/2$,
while the relative coordinate $x-y$ for two variables $x$ and $y$
are Wigner (or Fourier) transformed.
Namely, the $X$ dependence of the Wigner transform,
\begin{align}
A_w(X,p)=\int ds\, e^{ips} A(X+s/2,X-s/2)
\ ,
\end{align}
is assumed to be small.
The gradient expansion of the Wigner transform of the product is
given by using the Poisson bracket as,
\begin{align}
&\left[A B\right]_w(X,p)
\equiv \int ds e^{ips}\int dz\, A(X+\frac{s}{2},z)\, B(z,X-\frac{s}{2})
\nonumber\\
&=A_w(X,p) B_w(X,p)+\frac{i}{2}\PB{A_w,B_w}
 + {\cal O}(\frac{\partial^2}{\partial X^2})
 \label{gra}
\ ,\\
&\PB{A_w,B_w}\equiv
\frac{\partial A_w}{\partial p^\mu}
\frac{\partial B_w}{\partial X_\mu}
-
\frac{\partial A_w}{\partial X_\mu}
\frac{\partial B_w}{\partial p^\mu}\
\ .
\end{align}
(Here after we omit the subscript $w$.)

\begin{widetext}
We take the difference of Eqs.~(\ref{SDR}) and (\ref{SDL})
and carry out the Wigner transformation.
By assuming that $t_0$ is small enough ($t_0 \to -\infty$)
or $X^0$ is large enough,
we find
\begin{align}
&
\left [
  2i p\cdot \frac{\partial}{\partial X}
  +\frac{i}{2}\cdot \frac{\lambda(N+2)}{6N} 
  \int \!\! \frac{d^{d+1}k} { (2\pi) ^{d+1}}
  \left (
    \frac{\partial G^{aa}(X,k)}{\partial X}
   +\frac{\partial G^{bb}(X,k)}{\partial X}
  \right)
  \cdot \frac{\partial}{\partial p} 
\right ] G^{ab}  (X,p)
=iI^{ab}(X,p)
\label{eq:dif}
\\
&I^{ab}(X,p)
=\left[
   \Sigma^{ac}_\mathrm{nonl}c^{cd}G^{db}
-  G^{ac}c^{cd}\Sigma_\mathrm{nonl}G^{db}
\right]_w(X,p)
\nonumber\\
&~~~~~~~~~~~~~~~=\int \! ds e^{ip\cdot s}
 \int \!dz \left[
     \Sigma^{ac}_{\rm nonl}(X+s/2,z)c^{cd}G^{db}(z,X-s/2)
     -G^{ac}(X+s/2,z)c^{cd}\Sigma_{\rm nonl}^{db}(z,X-s/2)
	\right]
\label{eq:dif2}
\ ,
\end{align}
where 
$p$ is the momentum conjugate to $x-y$.
The gradient expansion with respect to the center-of-mass coordinate
$X$ is appropriate when the $X$-dependence of the system is smooth enough
(See for example~\cite{{BM},{IKV4}}).
The 1st order terms in the gradient expansion
are taken for the Green's functions and the self-energies.
We assume here that the interval of
$x^0 - y^0$ can take sufficiently large values for a fixed $X^0$.

By using Eq.~(\ref{gra}),
the gradient expansion of $I^{ab}$ for $(a,b)=(1,2)$ and $(2,1)$
is found to be,
\begin{align}
I^{12}(X,p)
=&
 (\Sigma^{11}_\mathrm{nonl}+\Sigma^{22}_\mathrm{nonl}) G^{12}
-(G^{11}+G^{22})\Sigma^{12}_\mathrm{nonl}
+\frac{i}{2}\PB{\Sigma^{11}_\mathrm{nonl}-\Sigma^{22}_\mathrm{nonl},G^{12}}
 -\frac{i}{2}\PB{G^{11}-G^{22},\Sigma^{12}_\mathrm{nonl}}
\nonumber\\
=&-C(X,p)+\PB{\mathrm{Re}\Sigma_R,G^{12}}
-\PB{\mathrm{Re}G_R,\Sigma_\mathrm{nonl}^{12}}
\ ,
\label{eq:I12}
\\
I^{21}(X,p)
=&-C(X,p)+\PB{\mathrm{Re}\Sigma_R,G^{21}}
-\PB{\mathrm{Re}G_R,\Sigma_\mathrm{nonl}^{21}}
\ ,
\end{align}
\end{widetext}
where the term $C$,
\begin{align}
C(X,p)=i \left[
       \Sigma_\rho(X,p)\; F(X,p) - \Sigma_F(X,p)\;  \rho(X,p)
        \right]
\label{Eq:coll}
\end{align}
may be identified as the collision term in the Boltzmann limit. 
In the last line of Eq.~(\ref{eq:I12}),
we have utilized the following relations,
\begin{align}
G^{12,21}=F \pm \frac{i}{2} \rho
\ ,\quad
&\Sigma_\mathrm{nonl}^{12,21}=\Sigma_F \pm \frac{i}{2} \Sigma_\rho
\ ,\\
G^{11,22}=F\mp i \mathrm{Re}G_R
\ ,\quad
&\Sigma_\mathrm{nonl}^{11,22}=\Sigma_F\mp i \mathrm{Re}\Sigma_R
\label{eq:1122}
\ .
\end{align}
The equation of motion for $G^\alpha (\alpha=12,21)$ are then rewritten as
\begin{align}
\PB{h,G^{12}}&+\PB{\mathrm{Re}G_R,\Sigma^{12}_\mathrm{nonl}}=-C
\label{EOM12}
\ ,\\
\PB{h,G^{21}}&+\PB{\mathrm{Re}G_R,\Sigma^{21}_\mathrm{nonl}}=-C
\label{EOM21}
\ ,\\
h(X,p)=&p^2 - m^2
- \frac{\lambda(N+2)}{6N}\int \frac{d^{d+1}k}{(2\pi)^{d+1}} F(X,k)
\nonumber\\
&- \mathrm{Re}\Sigma_R(X,p)\ .
\end{align}
Here we notice that the leading order of the 
gradient expansion of $C(X,p)$
is already 1st order $\sim O(\partial/\partial X)$~\cite{IKV4,Kita}.

Next let us take the difference of 
Eq. (\ref{EOM12})  
multiplied by $\ln (iG^{12}/\rho)$
and 
Eq. (\ref{EOM21})  
multiplied by $\ln (iG^{21}/\rho)$.
Then we integrate the resultant expression 
over $d^{d+1} p/ (2\pi)^{d+1}$ to arrive
at the following equation:
\begin{eqnarray}
\partial _\mu s^{\mu} 
=
\frac{1}{2} \int \!\! 
\frac{d^{d+1}p}{(2\pi)^{d+1}} \ln \frac{G^{12}}{G^{21}} C(X,p) .
\label{entropy}
\end{eqnarray} 
In Eq.~(\ref{entropy}), we have  defined the entropy current $s^\mu (X)$ as
\begin{align}
s^\mu 
=
 \int \!\! \frac{d^{d+1}p}{(2\pi)^{d+1} } 
&\left [
\left(
   p^\mu -\frac{1}{2}\frac{\partial {\rm Re} \Sigma_R}{\partial p_\mu} 
\right)
\right .
\nonumber\\
\times&\left ( 
   -G^{12}\ln \frac{iG^{12}}{\rho}+G^{21} \ln \frac{i G^{21}}{\rho}
\right) 
\nonumber \\
-\frac{1}{2} {\rm Re} G_R \
&
\left .
\left(
   -\frac{\partial}{\partial p_\mu} 
    \left(  \frac{ \Sigma_\rho}{i} \frac{iG^{12}}{\rho}  \right)
\ln \frac{iG^{12}}{\rho} 
\right.
\right.
\nonumber\\
&\left.
\left.
   + \frac{\partial}{\partial p_\mu} 
     \left (  \frac{\Sigma_\rho}{i} \frac{iG^{21}}{\rho}
     \right )
\ln \frac{iG^{21}}{\rho}  
\right ) 
\right  ]
\, ,
\end{align}
where 
we have used the relations 
$i(\Sigma^{11}-\Sigma^{22})=2{\rm Re}\Sigma_R$
and  $i(G^{11}-G^{22})=2{\rm Re} G_R$
from Eq.~(\ref{eq:1122}).
We have also adopted the approximations,
\begin{align}
\Sigma^{12} \simeq \Sigma_\rho G^{12}/\rho
\ ,\quad
\Sigma^{21} \simeq \Sigma_\rho G^{21}/\rho
\ ,
\end{align}
in the 1st order gradient expansion~\cite{IKV4,Kita,Cassing}.

In the end we shall write the two-point functions
in the form of the Kadanoff-Baym Ansatz
$G^{12}=-i\rho f$ and $G^{21}=-i\rho (1+f)$ with
a real function $f$,
then the above entropy current can be rewritten as
\begin{align}
s^\mu = \int \!\! \frac{d^{d+1}p}{(2\pi)^{d+1} } 
 \left[ \frac{\rho}{i} \left( p^\mu-\frac{1}{2} \frac{\partial {\rm Re} \Sigma_R}{ \partial p_\mu} \right) 
+ \frac{ \Sigma_\rho}{i} \frac{1}{2}\frac { \partial {\rm Re} G_R}{\partial p_\mu} \right] \sigma
\; ,
\label{s3}
\end{align}
where we introduced the notation
\begin{eqnarray}
\sigma(X,p) = -f \ln f +(1+f) \ln (1+f) \; .
\end{eqnarray}
In the quasiparticle limit $\Sigma_{\rm nonl} \rightarrow 0$,
we know that the Green's function is given by the occupation number $n_{\bf p}$,
\begin{align}
G^{12}=&-i\rho f= 2\pi \delta ((p^0)^2-\Omega_{\bf p}^2)(\theta(-p^0)+n_{\bf p})
\ ,\\
G^{21}=&-i\rho (1+f)= 2\pi \delta ((p^0)^2-\Omega_{\bf p}^2)(\theta(p^0)+n_{\bf p})
\ ,\\
\Omega_{\bf p}^2=&{\bf p}^2+m^2+ \frac{\lambda(N+2)}{6N}
\int \frac{d^dk}{ (2\pi)^d}\, F
\ .
\end{align}
Then we reproduce the normal entropy current for bosons
\begin{align}
s^\mu \rightarrow \int \frac{d^dp}{ (2\pi)^d} \frac{p^\mu}{\epsilon_p}
\left[
-n_{\bf p} \ln n_{\bf p}
+(1+n_{\bf p})\ln (1+n_{\bf p})
\right]
\ .
\end{align}
The remaining work is to prove that the R.H.S.~of Eq.~(\ref{entropy}) becomes positive definite.

\subsection{NLO self-energy of $1/N$ expansion and H-theorem}
In this subsection we prove the H-theorem for the NLO self-energy
in the $O(N)$ model by use of the gradient expansion
of the \KB\ equation.
In the previous section and Ref.~\cite{Nishiyama},
we find the relation (\ref{entropy})
where the divergence of entropy density is proportional to
$C=i(\Sigma_\rho F- \Sigma_F \rho)$ for the scalar field theory. 
We shall Wigner (or Fourier) transform
$\hat \Sigma_F$ and $\hat \Sigma_\rho$
in addition to $\hat D_F$ and $\hat D_\rho$
with $\hat \Pi_F$ and $\hat \Pi_\rho$.

\begin{widetext}
First in order to derive $\hat D_\rho(X,p)$ we calculate $\hat D_R(X,p)$ and $\hat D_A(X,p)$ by multiplying 
$\theta(x^0-y^0)$ and $- \theta (y^0-x^0)$ in (\ref{hdrxy}) and
Wigner transform. 
Then $\hat D_R(X,p)$ can be calculated as,
\begin{align} 
\hat D_R(x,y) \equiv& \theta(x^0-y^0) \hat D_\rho(x,y)
\nonumber\\
=&-\frac{\lambda}{3N} \theta(x^0-y^0) \hat \Pi_\rho(x,y)
+ \frac{\lambda}{3N}\theta(x^0-y^0) \int^{x^0}_{y^0}dz\  \hat \Pi_\rho(x,z) \hat D_\rho(z,y)
\nonumber \\
=& -\frac{\lambda}{3N} \theta(x^0-y^0) \hat \Pi_\rho(x,y)
+ \frac{\lambda}{3N}\theta(x^0-y^0) \int^{\infty}_{t_0}dz\  \hat \Pi_\rho(x,z) \hat D_\rho(z,y) 
(\theta(x^0-z^0)- \theta(y^0-z^0))
\nonumber \\
&\left({\rm due \ to} \ \theta(x^0-y^0)\theta (x^0-z^0)
= \theta(y^0-z^0)\theta(x^0-y^0) + \theta(z^0-y^0)\theta (x^0-z^0) \right)
\nonumber \\
=& -\frac{\lambda}{3N} \theta(x^0-y^0) \hat \Pi_\rho(x,y)
+ \frac{\lambda}{3N} \int^{\infty}_{t_0}dz\  \hat \Pi_\rho(x,z) \hat D_\rho(z,y)  \theta(z^0-y^0)\theta(x^0-z^0)
\nonumber \\
=& -\frac{\lambda}{3N} \hat \Pi_R(x,y)
+ \frac{\lambda}{3N} \int^{\infty}_{t_0}dz\  \hat \Pi_R(x,z) \hat D_R(z,y) 
\ .
\end{align}
\end{widetext}
By taking $t_0\rightarrow -\infty$, Wigner transformation,
and using the gradient expansion relation (\ref{gra}),
$\hat D_R(X,p)$ can be written in the leading order of the gradient expansion as
\begin{eqnarray}
\hat D_R(X,p)= - \frac{\lambda}{3N} 
(\hat \Pi_R(X,p) - \hat \Pi_R(X,p)\hat D_R(X,p))
\ .
\end{eqnarray}
As a result we obtain $\hat D_R(X,p)$ can be derived as
\begin{eqnarray} 
\hat D_R(X,p)= \frac{- \frac{\lambda}{3N} \hat \Pi_{R}(X,p)}{1- \frac{\lambda}{3N} \hat \Pi_R(X,p)} 
\label{hdrp}
\end{eqnarray}
where we have taken only the leading order (local parts)
of the gradient expansion in the convolution by use of (\ref{gra}).
\footnote{Here it might be necesarry to extract the nonlocal parts in estimating memory correction
terms although it is not assured that the nonlocal parts is within 1st order 
gradient expansion.}
In a similar way $D_A(X,p)= -{\rm F.T.} \theta(y^0-x^0)\hat D_\rho(x,y)$ can be written as
\begin{eqnarray} 
\hat D_A(X,p)= - \frac{\lambda}{3N}
\left[\hat \Pi_A(X,p)-\hat \Pi_A (X,p)\hat D_A (X,p)\right]
\ ,
\end{eqnarray}
where we have used the relation $\theta(y^0-x^0) \theta(y^0-z^0)= \theta(x^0-z^0)\theta(y^0-x^0) + \theta(z^0-x^0)
\theta(y^0-z^0)$.
As a result we obtain 
\begin{eqnarray} 
\hat D_A(X,p) = \frac{- \frac{\lambda}{3N} \hat \Pi_A(X,p)}{ 1- \frac{\lambda}{3N}\hat \Pi_A(X,p)}.
\label{hdap}
\end{eqnarray}
\begin{widetext}
By using the relations $\hat D_\rho(X,p)= \hat D_R(X,p)-\hat D_A(X,p)$, (\ref{hdrp}) and (\ref{hdap}),
we obtain
\begin{align} 
\hat D_\rho(X,p)=&\hat D_R(X,p) -\hat D_A(X,p)
= \frac{-\frac{\lambda}{3N} \hat \Pi_\rho(X,p)}
{ \left(1-\frac{\lambda}{3N}\hat \Pi_R(X,p) 
\right)
 \left(1- \frac{\lambda}{3N}\hat \Pi_A(X,p) \right) }
= - \frac{\lambda_{\rm eff}(X,p)}{3N} \hat \Pi_\rho(X,p),
\label{hdrxp}
\end{align}
where we have defined the effective coupling as,
\begin{eqnarray}
\lambda_{\rm eff}(X,p)= \frac{\lambda}{ \left(1-\frac{\lambda}{3N}\hat \Pi_R(X,p) 
\right)
 \left(1- \frac{\lambda}{3N}\hat \Pi_A(X,p) \right) }. 
\end{eqnarray}

In a similar way $\hat D_F(x,y)$ can be derived by rewriting as 
\begin{align} 
\hat D_F(x,y)=& - \frac{\lambda}{3N}\hat \Pi_F(x,y)
- \frac{\lambda}{3N} \int_{t_0}^{y^0} dz \ 
\left(\hat D_F(x,z)  \hat \Pi_\rho(z,y)
- \hat D_\rho(x,z)  \hat \Pi_F(z,y)\right)
\nonumber \\
=& - \frac{\lambda}{3N}\hat \Pi_F(x,y) 
+ \frac{\lambda}{3N} \int_{t_0}^{\infty} dz \ 
\left(\hat D_F(x,z) \hat \Pi_A(z,y)
+\hat D_R(x,z)  \hat \Pi_F(z,y)\right)
\ .
\end{align}
Then the Wigner transform $\hat D_F(X,p)$ is derived
by taking $t_0\rightarrow -\infty$ as
\begin{align}
&\hat D_F(X,p) = - \frac{\lambda}{3N} \hat \Pi_F
+ \frac{\lambda}{3N} \left(\hat D_F \hat \Pi_A
+  \hat D_R\hat \Pi_F \right) 
=\frac{- \frac{\lambda}{3N}\hat \Pi_F}{\left(1- \frac{\lambda}{3N}\hat \Pi_R \right) 
\left(1- \frac{\lambda}{3N}\hat \Pi_A \right)}
= - \frac{\lambda_{\rm eff}(X,p)}{3N} \hat \Pi_F(X,p)
\ ,
\label{hdfxp} 
\end{align}
where we have used Eq. (\ref{hdrp}). 
By use of (\ref{hdrxp}) and (\ref{hdfxp}) we obtain the following relations with respect to Wigner
transformed $\hat \Sigma_F(X,p)$ and $\hat \Sigma_\rho(X,p)$ from (\ref{sigfab}) and (\ref{sigrab}) as
\begin{align}
\hat \Sigma_F(X,p)
=&
 \int dq \left(F(X,q)\hat D_F(X,p-q) - \frac{1}{4} \rho(X,q) 
\hat D_\rho(X,p-q) 
 \right)
\nonumber \\
=& 
 \int dq \left(- \frac{\lambda_{\rm eff}(X,p-q)}{3N} \right) 
\left(F(X,q)\hat \Pi_F(X,p-q) - \frac{1}{4} \rho(X,q) \hat \Pi_\rho(X,p-q) \right)
\ ,\\
\hat \Sigma_\rho (X,p)
=&
\int dq \left( \rho(X,q) \hat D_F(X,p-q) + F(X,q)\hat D_\rho(X,p-q)  \right)
\nonumber \\
=&
 \int dq \left(- \frac{\lambda_{\rm eff}(X,p-q)}{3N} \right)
\left( \rho(X,q) \hat \Pi_F(X,p-q) + F(X,q) \hat \Pi_\rho(X,p-q) \right),
\end{align}
where $\hat \Pi_F(X,p)$  and $\hat \Pi_\rho(X,p)$ can be expressed from (\ref{pifxy}) and (\ref{pirxy}) as
\begin{align}
\hat \Pi_F(X,p) =&   \frac{\lambda}{6} \int dq \left( F(X,p-q) F(X,q) - \frac{1}{4} \rho(X,p-q) 
\rho(X,q) \right)
\label{pifxp}
\ ,\\
\hat \Pi_\rho(X,p) =& \frac{\lambda}{3} \int dq \ F(X,p-q) \rho(X,q)
\ .
\label{pirxp} 
\end{align}

In the end $\hat \Sigma_F(X,p)$ and $\hat \Sigma_\rho(X,p)$ 
can be reexpressed by use of (\ref{pifxp}) and (\ref{pirxp}) as
\begin{align}
\hat \Sigma_F(X,p)
=&  \int dq
	\left(
	F(X,q) \hat D_F(X,p-q) - \frac{1}{4} \rho(X,q) \hat D_\rho(X,p-q)
	\right) 
\nonumber \\
=& - \frac{\lambda}{6} \int dq dl \frac{\lambda_{\rm eff}(X,p-q)}{3N}  \Big[ F(X,q) F(X,l) F(X,p-q-l)
\nonumber \\
&- \frac{1}{4}
F(X,q) \rho(X,l) \rho(X,p-q-l) - \frac{1}{2} \rho(X,q) F(X,l) \rho(X,p-q-l) \Big]
\label{eq:sigfxp2}
\ ,\\
\hat \Sigma_\rho(X,p)
=& \int dq dl \left [
		\rho(X,q) \hat D_F(X,p-q) + F(X,q) \hat D_\rho(X,p-q)
		\right]
\nonumber \\
=& - \frac{\lambda}{6} \int dq dl
	\frac{\lambda_{\rm eff}(X,p-q)}{ 3N}
	\Big[ \rho(X,q) F(X,l) F(X,p-q-l)
\nonumber \\
&
- \frac{1}{4} \rho(X,q)\rho(X,l) \rho(X,p-q-l) + 2 F(X,q) F(X,l) \rho(X,p-q-l) \Big]
\label{eq:sigrxp2}
\end{align}

Finally 
we obtain the following collision term $C(X,p)=i(\hat \Sigma_\rho (X,p)F(X,p)- \Sigma_F(X,p) \rho(X,p))$ in Eq. (\ref{Eq:coll}),
\begin{align}
C(X,p_1)
=&\frac{\lambda}{18N} 
\int dp_2 dp_3 dp_4 \  \delta^{d+1} (p_1+p_2-p_3-p_4) \lambda_{\rm eff}(X,p+p_2) 
K(X,p_1,p_2,p_3,p_4)
\label{Eq:collB}
\ ,\\
K(X,p_1,p_2,p_3,p_4)
\equiv&
  G^{12}(X,p_1) G^{12}(X,p_2) G^{21}(X,p_3) G^{21}(X,p_4)
- G^{21}(X,p_1) G^{21}(X,p_2) G^{12}(X,p_3) G^{12}(X,p_4)
\ ,
\end{align}
where we have used the self-energy
(\ref{eq:sigfxp2}), (\ref{eq:sigrxp2}),
$F(X,-q)=F(X,q)$, $\rho(X,-q)=-\rho(X,q)$
and
\begin{eqnarray}
F(X,l)F(X,r)-\frac{1}{4} \rho(X,l) \rho(X,r)
&=& \frac{1}{2} \left[ G^{12}(X,l) G^{12}(X,r) + G^{21}(X,l) G^{21}(X,r)\right]
\ ,
\nonumber \\
\rho(X,l)F(X,r) + F(X,l) \rho(X,r)
&=& i \left[ G^{21}(X,l) G^{21}(X,r)-G^{12}(X,l) G^{12}(X,r) \right]
\ ,
\end{eqnarray}
with the relations $F(X,p)=\frac{1}{2}\left[G^{21}(X,p)+G^{12}(X,p) \right]$ and 
$\rho(X,p)=i\left[G^{21}(X,p)-G^{12}(X,p)\right]$.

As a result,
the divergence of the entropy current per particle component can be written as
\begin{align}
&\partial_\mu s^\mu (X)
= \frac{1}{2} \int dp \left( \ln\frac{G^{12}(X,p)}{G^{21}(X,p)} \right) 
i \left(\hat \Sigma_\rho(X,p) F(X,p) - \hat \Sigma_F(X,p)\rho(X,p) \right)
\nonumber \\
=& \frac{\lambda}{36N} \int d\Gamma\,
K(X,p_1,p_2,p_3,p_4)\,
\ln \left[\frac{G^{12}(X,p_1)}{G^{21}(X,p_1)} \right]
\nonumber \\
=& \frac{\lambda}{36N} \cdot \frac12 \int d\Gamma\,
K(X,p_1,p_2,p_3,p_4)\,
\ln \left[\frac{G^{12}(X,p_1)G^{12}(X,p_2)}{G^{21}(X,p_1)G^{21}(X,p_2)} \right]
\nonumber \\
=& \frac{\lambda}{36N} \cdot \frac14 \int d\Gamma\,
K(X,p_1,p_2,p_3,p_4)\,
\ln \left[\frac{G^{12}(X,p_1)G^{12}(X,p_2)G^{21}(X,p_3)G^{21}(X,p_4)}
{G^{21}(X,p_1)G^{21}(X,p_2)G^{12}(X,p_3)G^{12}(X,p_4)}
\right]
\nonumber \\
=& \frac{\lambda}{36N} \cdot \frac{1}{4} \int d\Gamma\,
\left[G^{12}(X,p_1)G^{12}(X,p_2) G^{21}(X,p_3) G^{21}(X,p_4)
- G^{21}(X,p_1)G^{21}(X,p_2)G^{12}(X,p_3) G^{12}(X,p_4) \right]
\nonumber \\
&\times 
\ln \left(
\frac{G^{12}(X,p_1)G^{12}(X,p_2)G^{21}(X,p_3)G^{21}(X,p_4)}
     {G^{21}(X,p_1)G^{21}(X,p_2)G^{12}(X,p_3)G^{12}(X,p_4)} \right)
\geq 0
\label{eq:Htheorem}
\ ,\\
&d\Gamma\equiv
\prod_{i=1}^4 dp_i \delta^{d+1}(p_1+p_2-p_3-p_4) \lambda_{\rm eff}(p+q)
\ .
\end{align}
\end{widetext}
In this derivation, we have utilized
the inequality $(x-y)\ln(x/y)\geq 0$,
and the symmetry of $K$
under the exchange of momenta,
$p_1 \leftrightarrow p_2$ and
$(p_1,p_2) \leftrightarrow (p_3,p_4)$,
\begin{align}
K(X,p_1,p_2,p_3,p_4)=&K(X,p_2,p_1,p_3,p_4)
\nonumber\\
=&-K(X,p_3,p_4,p_1,p_2)
\ .
\end{align}
Hence we have proved the H-theorem for the NLO of $1/N$ expansion in the range of 1st order gradient expansion.
We can see that
any change of Green's functions with nonzero collision term 
$C= i\left[\hat \Sigma_\rho F- \hat \Sigma_F \rho \right] \not = 0$
contributes to entropy production.

The equality in Eq.~(\ref{eq:Htheorem}) holds when $K=0$ is satisfied
for all combinations of momenta which satisfy the energy-momentum conservation.
This condition is realized when $G^{21}/G^{12}$ is a linear function 
of $p$,
\begin{eqnarray}
\ln \frac{G^{21}(X,p)}{G^{12}(X,p)} =\ln \frac{1+f(X,p)}{f(X,p)} = \alpha(X) + \beta^\mu (X) p_\mu
\end{eqnarray}
where $\alpha(X)$ and $\beta(X)$ are arbitrary functions for the center of coordinate $X$.
Thus we notice that  at equilibrium the famous local equilibrium distribution function
\begin{eqnarray}
f(X,p)= \frac{1}{e^{\alpha(X)+\beta^\mu(X)  p_\mu}- 1}
\end{eqnarray}
is realized.
This is one of our new results.

\begin{figure}[tb]
\begin{center}
\includegraphics[width=4cm]{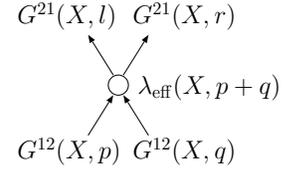}
\caption{
Collision term $C$
}
\label{fig:coll}
\end{center}
\end{figure}

Equation~(\ref{Eq:collB}) shows that $C$ actually has
the form of the collision term of bosons.
The first term in $K$ represents the loss term,
which is proportional to the product of probability
$G^{12}(p_1)G^{12}(p_2)\propto f(p_1)f(p_2)$
for particles having $p_1$ and $p_2$,
and proportional to the bosonic enhancement factor in the final state,
$G^{21}(p_3)G^{21}(p_4)\propto (1+f(p_3))(1+f(p_4))$,
as schematically shown in Fig.~\ref{fig:coll}.

\section{Entropy from thermodynamic potential in thermal equilibrium}
In this section we derive an analytic expression of the entropy density
in thermal equilibrium from the thermodynamic potential.
Its derivation is based on Refs.~\cite{BIR1999,BIR2001}
and we adopt the imaginary time formalism. 
The new point in this article is the derivation of entropy density
in the presence of NLO skeletons of the $1/N$ expansion at equilibrium.  
At thermal equilibrium, the entropy density is given by differentiating
the 2PI effective action or thermodynamic potential 
with respect to temperature $T$.  
We find that the resultant entropy in the $O(N)$ theory corresponds to 
the kinetic entropy (\ref{s3}) without memory corrections
in the limit of thermal equilibrium
for NLO skeletons of the $1/N$ expansion. 

First let us begin with the Lagrangian
\begin{align}
{\cal L}[ \phi]
= \frac{1}{2} \partial  \phi_a \partial \phi_a 
- \frac{1}{2}m^2  \phi_a^2
- \frac{\lambda}{4!N} \left( \phi_a  \phi_a \right)^2
\ ,
\end{align}
where $a=1 \sim N$.
The thermodynamic potential $\Omega= -P V$, which corresponds
to the 2PI effective action, of the scalar field can be written
as the following functional of the full Green's function
$G$~\cite{{LW},{CJT}}:
\begin{align}
\beta\Omega[G]/N
=& - \frac{1}{N} {\rm log} Z
\nonumber\\
=& \frac{1}{2} {\rm Tr }{\rm log} G^{-1} - \frac{1}{2} {\rm Tr} \Sigma G
  + \frac{1}{2} \Phi[G]/N
\ ,
\end{align}
where $Z$ represents the generating functional,
${\rm Tr}$ denotes the trace in configuration space,
$\beta=1/T$,
and $\Phi[G]/2$ is the infinite series of 2PI skeletons.
Here we have expressed the functional per particle components $N$.
The self-energy $\Sigma$ is defined by 
\begin{eqnarray}
\Sigma \equiv  \frac{\delta \Phi[G]}{\delta G}\ . 
\end{eqnarray}
The self-energy is related to the Green's function $G$
by the Schwinger-Dyson equation $ G^{-1} = G_0^{-1} + \Sigma$,
where $G_0^{-1}$ is the bare Green's function.
The Schwinger-Dyson equation is obtained
by imposing the stationary condition of the thermodynamic potential $\Omega$: 
\begin{eqnarray}
\delta \Omega[G]/ \delta G =0.
\label{eq:TPSP}
\end{eqnarray}
We discuss thermodynamic potential only under the stationary condition
in the following analyses.

The thermodynamic potential $\Omega[G]$ can be given in the momentum space as
\begin{align}
\frac{\Omega}{NV}
= \int \frac{d^{d+1} k}{(2\pi)^{d+1}} f(k^0)\,
{\rm Im}\, \left[\log (G^{-1}) - \Sigma G \right]
+ \frac{T \Phi[G]}{2NV}
\ ,
\label{eq:TP}
\end{align}
where $d+1$ is the space-time dimension and $f(k^0)= {1}/({e^{\beta k^0}-1})$.
The summation with Matsubara frequencies (Matsubara sum)
can be converted to the integral over $k^0$
of the integrand multiplied by the Bose function $f(k^0)$
by using standard contour integration techniques.
The Green's function $G(k)$ is written analytically
in terms of the spectral function;
\begin{eqnarray}
G(k^0, {\bf k})= \int _{- \infty} ^{\infty} \frac{dq^0}{(2\pi)}
\frac{\rho(q^0,{\bf k})}{q^0- k^0}
\ .
\label{eq:D}
\end{eqnarray}
For real variable $k^0$, we define the regularization of the integral as,
\begin{align}
{\rm Im} G(k^0,{\bf k})
\equiv&  {\rm Im} G(k^0+ i \epsilon, {\bf k}) = \frac{\rho(k^0,{\bf k})}{2}\ ,
\label{eq:Drho}
\\
{\rm Re} G(k^0,{\bf k})
\equiv&  {\rm Re} G(k^0+ i \epsilon, {\bf k}) = {\rm Re} G_R(k^0,{\bf k})\ .
\label{eq:Dre}
\end{align}
The regularization for the self-energy $\Sigma$
is defined in a similar way.

Next we shall derive the entropy density $\cal S$
by differentiating $\Omega$ by $T$:
\begin{eqnarray}
{\cal S} = - \frac{\partial \Omega/V}{ \partial T}.
\end{eqnarray}
The thermodynamic potential, as given in Eq.~(\ref{eq:TP}),
depends on its temperature through the distribution function $f(k^0)$
and the spectral function $\rho$ (or the Green's function $G$). 
By use of Eq.~(\ref{eq:TPSP}), temperature derivative of the Green's function or 
spectral function $\rho$ cancels out in the entropy density due to SD equation,
and we obtain~\cite{{BIR1999},{BIR2001},Riedel1968,VB1998}
\begin{align}
{\cal S} =& - N\int \frac{d^{d+1} k}{(2\pi)^{d+1}} \frac{\partial f(k^0)}{\partial T} {\rm Im} \ {\rm log} G^{-1}(k) 
\nonumber\\
&+ N \int \frac{d^{d+1} k}{(2\pi)^{d+1}} \frac{\partial f(k^0)}{\partial T} {\rm Im} \Sigma (k) {\rm Re} G(k) + {\cal S'}
\end{align}
with
\begin{align}
{\cal S'} = - \frac{\partial(T \Phi /V)}{\partial T} \Bigg|_G
+ N\int \frac{d^{d+1} k}{(2\pi)^{d+1}} \frac{\partial f(k^0)}{\partial T} {\rm Re} \Sigma (k) {\rm Im} G(k).
\label{eq:sprime}
\end{align}
By using the relation
\begin{align}
\frac{\partial f(k^0)}{\partial T}
=&- \frac{\partial f(k^0)}{ \partial k^0} {\rm log} \frac{1+f}{f}
= - \frac{\partial \sigma(k^0)}{\partial k^0}
\ ,\\
\sigma(k^0)
=&-f {\rm log} f + (1+f) {\rm log}(1+f)
\ ,
\end{align}
we can partially integrate with respect to $k^0$ and obtain
\begin{align}
&{\cal S}-{\cal S'}
\nonumber\\
=&  N
 \int \frac{d^{d+1} k}{(2\pi)^{d+1}} \left[ -{\rm Im} \left(G \frac{\partial G^{-1}}{\partial k^0}  \right)
+ \frac{\partial }{\partial k^0} ({\rm Im} \Sigma \ {\rm Re}G)  
  \right]\sigma
\nonumber \\
=&  N \int \frac{d^{d+1} k}{(2\pi)^{d+1}} \left[  \left(2 k^0  -  \frac{\partial {\rm Re}\Sigma}{\partial k^0} \right)
{\rm Im}G + {\rm Im} \Sigma \frac{\partial {\rm Re} G}{\partial k^0}
  \right] \sigma
\nonumber \\
=&  N \int \frac{d^{d+1} k}{(2\pi)^{d+1}} \left[
  \left(k^0  - \frac{1}{2} \frac{\partial {\rm Re}\Sigma_R}{\partial k^0} \right)
\rho
 + \frac{1}{2}   \frac{\partial {\rm Re} G_R}{\partial k^0}\Sigma_\rho 
  \right]\sigma
\end{align}
where we have used $G^{-1}=G_0^{-1}+ \Sigma= -k^2+m^2 +\Sigma$ and the definitions (\ref{eq:Drho}) and (\ref{eq:Dre}).
The remaining part is $\cal S'$ in the above expression.

In the end we shall examine whether 
\begin{eqnarray}
{\cal S'} = 0
\end{eqnarray}
is satisfied for both LO and NLO skeletons of the $1/N$ expansion.

It is easy to verify ${\cal S'}=0$ for LO skeleton of the $1/N$ expansion.
The LO skeleton is given by the expression
\begin{align}
&-\frac{T}{2V} \Phi^{({\rm LO})}
= - \frac{\lambda N}{24} 
\left(T \sum_\omega \int\! \frac{d^d k}{(2 \pi)^d} G(\omega,{\bf k})
 \right)^2
\nonumber \\
=& - \frac{\lambda N}{24}  
\int\! \frac{d^{d+1}k_1}{(2\pi)^{d+1}} 
\int\! \frac{d^{d+1}k_2}{(2\pi)^{d+1}}  \rho(k_1) \rho(k_2) f(k_1^0) f(k_2^0)
\end{align}
where 
we have used Eq.~(\ref{eq:D}) and the relation
\begin{eqnarray}
T\sum _{\omega=i2\pi n T} \frac{1}{k^0-\omega}= f(k^0).
\label{eq:omega-f}
\end{eqnarray}
Then we take account of the temperature dependence
of the occupation number function $f$.
By differentiating by $T$ we obtain the first component in $\cal S'$ (\ref{eq:sprime}) for
the LO skeleton as
\begin{align}
&-\frac{\partial }{\partial T}
	\left(\frac{T}{2V} \Phi^{({\rm LO})} \right) \Bigg|_G
\nonumber\\
=& 
-\frac{\lambda N}{12}
 \int \! \frac{d^{d+1}k_1}{(2\pi)^{d+1}} 
 \int \! \frac{d^{d+1}k_2}{(2\pi)^{d+1}} f(k_1^0) \frac{\partial f(k_2^0)}{\partial T} \rho(k_1) \rho(k_2).
\label{eq:Phil1}
\end{align}
The second component of $\cal S'$ (\ref{eq:sprime}) has the real part of the self-energy which is given by opening
one line of the skeleton.
The real part of the self-energy $\Sigma =\frac{\delta \Phi[G]}{\delta G}$ is written as
\begin{eqnarray}
{\rm Re}\ \delta_{ab} \hat \Sigma^{({\rm LO})} = \delta_{ab} \hat \Sigma^{({\rm LO})}= \delta_{ab} \frac{\lambda}{6}   
\int \! \frac{d^{d+1}k}{(2\pi)^{d+1}} f(k^0) \rho(k).
\label{eq:real:l1}
\end{eqnarray}
This relation (\ref{eq:real:l1}) gives the following second component of $\cal S'$ (\ref{eq:sprime}):
\begin{align}
&N \int \! \frac{d^{d+1}q}{(2\pi)^{d+1}} \frac{\partial f(q^0)}{\partial T} {\rm Re}  \Sigma^{({\rm LO})} (q) 
{\rm Im} G(q)
\nonumber\\
=& 
\frac{\lambda N}{12} 
 \int \! \frac{d^{d+1}k}{(2\pi)^{d+1}} 
 \int \! \frac{d^{d+1}q}{(2\pi)^{d+1}} f(k^0) \frac{\partial f(q^0)}{\partial T} \rho(k) \rho(q),
\label{eq:PiDl1}
\end{align}
where we have used ${\rm Im}\ G=\rho/2$.
We can confirm that two components (\ref{eq:PiDl1}) and (\ref{eq:Phil1})
cancel each other precisely
and that ${\cal S'}=0$ is verified for LO skeleton of $1/N$ expansion. 

Furthermore we shall derive the expression of ${\cal S}'$ for 
the NLO skeletons of the $1/N$ expansion in the $O(N)$ theory
by writing down two components of $\cal S'$ explicitly 
and investigate whether ${\cal S}'=0$ is satisfied or not in this order. 
We find the $n+1$-loop $(n \geq 2)$ order NLO skeleton per particle component
is given by
\begin{align}
&-\frac{T}{2V} \Phi^{(n+1)}
\nonumber\\
=& \frac{1}{2n} \left(-\frac{\lambda}{6} \right)^n T \sum_{\omega} \int \! 
\frac{d^d q}{(2\pi)^d} \left(  \prod_{j=1}^n T\sum_{\omega_j} \int \! \frac{d^d k_j}{(2\pi)^d} \right)
\nonumber \\ 
& \times G(\omega, {\bf q}) G(\omega_1, {\bf k}_1)
\nonumber \\ 
& \times G (\omega + \omega_1-\omega_2, {\bf q+k_1-k_2}) G(\omega_2,{\bf k_2})
\nonumber \\
& \times \cdot \cdot \cdot
\nonumber \\
& \times G(\omega+ \omega_1 -\omega_n, {\bf q+k_1-k_n})G(\omega_n, {\bf k_n}).
\label{eq:Phi-n1}
\end{align}
For the above skeletons the explicit form of 
first and second component in $\cal S'$ is given in Appendix A. 
Then we find that it might be possible to prove ${\cal S}'=0$
for each loop order of NLO skeletons of the $1/N$ expansion. 
However we also notice that these skeletons might have singularities
at the points where energy denominator vanishes for $n\geq 3$.
Memory correction terms can appear when momentum integration
with these denominators can not be regularized~\cite{IKV4,CP1975}. 
Then we have to carefully estimate the singularities
and add the memory correction terms in the definition of entropy current.
Hence the possibility of the appearance of memory correction terms
is not completely denied.

\section{Summary}
In this article, we have discussed the entropy density and the H-theorem
in the Kadanoff-Baym equation for the $O(N)$ theory
with the next-to-leading order (NLO) self-energy of the $1/N$ expansion.
While many numerical analyses show thermalization in the Kadanoff-Baym equation
takes place,
it is desired to confirm thermalization analytically
for a given form of the self-energy.
Thus we have derived the relativistic kinetic entropy and confirmed that
the H-theorem is satisfied for local parts of the NLO self-energy.
We have adopted the gradient expansion
and taken account of the 1st order terms of the Kadanoff-Baym equation.
The gradient expansion is an appropriate approximation
in treating moderately changing systems in space-time.
Since we ignore higher order derivative terms,
spatial fluctuations in a small volume are assumed to vanish.
Thus the gradient expansion corresponds to a coarse graining procedure.
Time reversal invariance is violated by the gradient expansion
of the (time reversible) Kadanoff-Baym equation.
Then it is suggested that the KB dynamics obtains the time arrow
at the level of the Green's function,
and that the observed system at late time will reach the maximum entropy
state where local equilibrium is realized.
The reliability of the gradient expansion and the form of the self-energy
are important in the proof of the H-theorem.
From the proof of the H-theorem we find that any change of the Green's function
with nonzero collision term contributes to entropy production. 
The entropy production is ensured at the level of the Green's function
(No need of quasiparticle approximation).
When local thermal equilibrium is realized,
the collision term contribution vanishes and entropy ceases to increase.
The proof of H-theorem is consistent with the numerical analyses
of the $O(N)$ model.
Thus the proof might provide a criterion whether thermalization occurs or not
even in the cases when numerical analyses are difficult to perform.

We have also considered the entropy density $\cal S$
derived from thermodynamic potential,
and compared with the kinetic entropy density.
When we only take account of the principal value part
in the momentum integral for  $\cal S$,
$\cal S$ corresponds to the kinetic entropy density in thermal equilibrium.
However if singularities remain in the energy denominators,  
memory correction terms might appear.
Hence entropy current have the possibility to have additional terms
in its definition. 

\section*{ Acknowledgment}

He would like to thank 
Profs.\ T.~Matsui, H.~Fujii and K.~Itakura for 
fruitful discussions in  non-equilibrium statistical physics.
The research in this paper has been supported by JSPS research fellowships for
Young Scientists under the grant number 21$\cdot$6697,
the Global COE Program
"The Next Generation of Physics, Spun from Universality and Emergence",
and the Yukawa International Program for Quark-hadron Sciences (YIPQS).

\begin{widetext}
\begin{appendix}

\section{The n-loop order skeleton and $\cal S'$}

In this appendix we shall confirm ${\cal S}'=0$ for the NLO skeletons order by order in
the coupling expansion.
The $n+1$ loop order component of $\Phi$ in $1/N$  expansion is given in Eq.~(\ref{eq:Phi-n1}).
This equation is transformed as 
\begin{eqnarray}
-\frac{T}{2V } \Phi^{(n+1)} = \frac{1}{2n} \left(- \frac{\lambda}{6} \right)^n T^{n+1} \sum_{\omega} \sum_{\omega_1}
\int \! \frac{d^{d+1} k_1}{(2\pi)^{d+1}} 
\int \! \frac{d^{d+1} q}{(2\pi)^{d+1}} 
\frac{\rho(q)}{q^0-\omega} \frac{\rho(k_1)}{k_1^0-\omega_1} 
\ \ \ \ \ \ \ \ \ \ \ \ \ \ 
\nonumber \\
\times   \left( \prod_{l=2}^n \sum_{\omega_l}
 \int \! \frac{d^{d+1} p_l}{(2\pi)^{d+1}} \int \! \frac{d^{d+1} k_l}{(2\pi)^{d+1}}
(2\pi)^d \delta^{(d)} ({\bf p_l +k_l -q -k_1})  \frac{\rho(p_l)}{p_l-(\omega+\omega_1-\omega_l)} 
\frac{\rho(k_l)}{k_l^0-\omega_l}\right)
\label{eq:Phi-n1-2}
\end{eqnarray}
with Eq.~(\ref{eq:Drho}).
We shall sum up $\omega_j$ ($j=1,\cdot \cdot \cdot , n$) by use of the relations  
\begin{eqnarray}
\prod _{l=2}^n T \sum _{\omega_l} \frac{1}{p_l^0 -(\omega + \omega_1 -\omega_l)} \frac{1}{ k_l^0 -\omega_l}
&=&
\prod _{l=2}^n T\sum _{\omega_l} 
\left( \frac{1}{p_l^0 -(\omega + \omega_1 -\omega_l)} + \frac{1}{ k_l^0 -\omega_l}\right)
\frac{1}{p_l^0 +k_l^0 -\omega_1 -\omega}
\nonumber \\
&=& \prod _{l=2}^n 
\frac{1}{p_l^0 +k_l^0 -\omega_1 -\omega} (f(k_l^0) +f(p_l^0) +1) 
\end{eqnarray}
with (\ref{eq:omega-f}) and  $-f(-p_l^0)=f(p_l^0)+1$, and 
\begin{eqnarray}
\frac{1}{k_1^0-\omega_1} \prod_{l=2}^n \frac{1}{ p_l^0 +k_l^0 -\omega_1 -\omega} =
\frac{1}{k_1^0-\omega_1} \prod_{l=2}^n \frac{1}{ p_l^0 +k_l^0 -\omega -k_1^0}  \ \ \ \ \ \ \ \ \ \ \ \  {\space}
\nonumber \\
+ \sum_{j=2}^n \frac{1}{p_j^0+k_j^0 -\omega-\omega_1} \frac{1}{ k_1^0 -(p_j^0+k_j^0 -\omega)} 
\prod_{l=2, l\not =j}^n \frac{1}{ (p_l^0+k_l^0 -\omega)- (p_j^0 + k_j^0-\omega)}.
\end{eqnarray} 
Then we obtain the factor $S_\omega$ in (\ref{eq:Phi-n1-2}) as
\begin{eqnarray}
S_{\omega} &\equiv & T \sum_{\omega_1} \frac{1}{k_1^0 -\omega} \prod_{l=2}^n T \sum_{\omega_l} 
\frac{1}{p_l^0-(\omega +\omega_1 -\omega_l)} \frac{1}{k_l^0-\omega_l} 
\nonumber \\
&=& f(k_1^0) \prod_{l=2}^n \frac{f(k_l^0 + f(p_l^0) +1}{p_l^0+k_l^0-\omega-k_1^0} 
+ \sum_{j=2}^n \frac{f(p_j^0) f(k_j^0) }{k_1^0 -(p_j^0 +k_j^0 -\omega)} 
\prod_{l=2, l\not =j}^n \frac{f(k_l^0) + f(p_l^0) +1}{p_l^0 + k_l^0 -p_j^0 -k_j^0}
\end{eqnarray}
with $f(p_j^0 + k_j^0) (f(p_j^0)+ f(k_j^0) +1) = f(p_j^0) f(k_j^0)$.
In the end by use of the relation 
\begin{eqnarray}
T \sum_{\omega} \frac{1}{q^0-\omega} S_{\omega} &=& f(q^0) f(k_1^0) \prod_{l=2}^n 
\frac{f(p_l^0)+ f(k_l^0) +1}{p_l^0 + k_l^0 -k_1^0 -q^0 }
\nonumber \\
&&+ \sum_{j=2}^n f(p_j^0) f(k_j^0) \frac{f(q^0)+ f(k_1^0) +1}{q^0+ k_1^0 -p_j^0-k_j^0 } 
\prod_{l=2, l\not =j}^n  \frac{f(p_l^0) + f(k_l^0)+1}{p_l^0 + k_l^0 -p_j^0 -k_j^0}
\end{eqnarray}
with (\ref{eq:omega-f}),
\begin{eqnarray}
\frac{1}{q^0 -\omega} \prod_{l=2}^n \frac{1}{p_l^0 + k_l^0 -\omega -k_1^0} = 
\frac{1}{q^0-\omega} \prod_{l=2}^n \frac{1}{p_l^0 + k_l^0 -k_1^0 -q^0} \ \ \ \ \ \ \ \ \ \ \ \ \ \ \ \ \ \ \ \ \ 
\ \ \ \ \ \ \ \ \ \ 
\nonumber \\
+ \sum_{j=2}^n \frac{1}{p_j^0 +k_j^0 -k_1^0 -\omega} \frac{1}{q^0 -(p_j^0 +k_j^0 -k_1^0)} \prod_{l=2, l\not = j}^n
\frac{1}{(p_l^0 +k_l^0 -k_1^0)- (p_j^0+k_j^0 -k_1^0)}
\end{eqnarray}
and
\begin{eqnarray}
\frac{1}{q^0-\omega} \sum_{j=2}^n \frac{1}{k_1^0 -p_j^0 -k_j^0 + \omega} 
= \sum_{j=2}^n \left( \frac{1}{q^0 -\omega} - \frac{1}{p_j^0 +k_j^0 -k_1^0 -\omega} \right) 
\frac{1}{k_1^0 +q^0 -p_j^0 -k_j^0 },
\end{eqnarray}
we obtain
\begin{eqnarray}
-\frac{T}{2V} \Phi^{(n+1)} 
&=& \frac{1}{2n} \left(-\frac{\lambda}{6} \right)^n \Bigg[ \int \! \frac{d^{d+1}q}{(2\pi)^{d+1} }
\int \! \frac{d^{d+1}k_1}{(2\pi)^{d+1} } \rho(q)\rho(k_1) 
\nonumber \\
&& \times \left( \prod_{j=2}^n 
\int \! \frac{d^{d+1}k_j}{(2\pi)^{d+1} }\int \! \frac{d^{d+1}p_i}{(2\pi)^{d+1} }
(2\pi)^d \delta^{(d)} ({\bf p_j+p_j - q-k_1}) \rho(p_j) \rho(k_j) \right) \Bigg]
\nonumber \\
&& \times \Bigg[ f(q^0) f(k_1^0) \prod_{l=2}^{n}  \frac{f(k_l^0) +f(p_l^0) +1 }{p_l^0 +k_l^0 - k_1^0 -q^0}
\nonumber \\
&&
+ \sum_{j=2}^n f(p_j^0) f(k_j^0) \frac{f(k_1^0) +f(q^0)+1}{q^0+k_1^0 -p_j^0 -k_j^0} \prod_{l=2, l\not =j}^n 
\frac{ f(p_l^0) + f(k_l^0) +1 }{p_l^0 +k_l^0 - p_j^0 -k_j^0} \Bigg].
\end{eqnarray}
(Here for $n=2$ since $f(k_l^0)+f(p_l^0)+1=f(k_l^0)f(p_l^0)(e^{\frac{k_l^0+p_l^0}{T}}-1)$, 
by use of the interchange $(q,k_1)\leftrightarrow (p_2,k_2)$ the above terms in the bracket 
can be transformed as
\begin{eqnarray}
f(q^0)f(k_1^0)\frac{f(p_2^0)+f(k_2^0)+1}{ p_2^0+k_2^0-q^0-k_1^0 } \rightarrow \frac{1}{2} f(q^0)f(k_1^0) f(p_2^0)f(k_2^0)
\frac{e^{\frac{p_2^0+k_2^0}{T}} -e^{\frac{q^0+k_1^0}{T}} }{p_2^0+k_2^0-q^0-k_1^0 },  
\end{eqnarray}
so that the skeleton $\Phi^{(3)}$ is analytic function and the fraction with energy denominator can 
be written by its principal value. )

Hence the first component of $\cal S'$ is given by   
\begin{eqnarray}
\frac{\partial }{\partial T} \left( -\frac{T}{2V} \Phi^{(n+1)} \right)\Bigg|_G &=&
\left(-\frac{\lambda}{6} \right)^n \frac{\partial f(q^0)}{\partial T} 
\Bigg[ \int \! \frac{d^{d+1}q}{(2\pi)^{d+1} }
\int \! \frac{d^{d+1}k_1}{(2\pi)^{d+1} } \rho(q)\rho(k_1) 
\nonumber \\
&& \times \left( \prod_{j=2}^n 
\int \! \frac{d^{d+1}k_j}{(2\pi)^{d+1} }\int \! \frac{d^{d+1}p_i}{(2\pi)^{d+1} }
(2\pi)^d \delta^{(d)} ({\bf p_j+p_j - q-k_1}) \rho(p_j) \rho(k_j) \right) \Bigg]
\nonumber \\
&&  \times  \frac{\partial f(q^0)}{\partial q^0}
 \Bigg[  f(k_1^0) \prod_{l=2}^{n}  \frac{f(k_l^0) +f(p_l^2) +1 }{p_l^0 +k_l^0 - k_1^0 -q^0}
\nonumber \\
&&
+ \sum_{j=2}^n  \frac{ f(p_j^0) f(k_j^0) }{q^0+k_1^0 -p_j^0 -k_j^0} \prod_{l=2, l\not =j}^n 
\frac{ f(p_l^0) + f(k_l^0) +1 }{p_l^0 +k_l^0 - p_j^0 -k_j^0} \Bigg].
\label{eq:1st}
\end{eqnarray} 
The second component $\int \! \frac{d^{d+1} q}{(2\pi)^{d+1}} {\rm Re} \hat \Sigma^{(n+1)}(q)  {\rm Im} G 
\frac{\partial f(q^0)}{\partial T}$ cancels the principal value of (\ref{eq:1st})
where ${\rm Im} G =\frac{\rho}{2}$ and
\begin{eqnarray}
\delta_{ab}\hat \Sigma^{(n+1)} (\omega, {\bf q}) & = & \frac{\delta \Phi^{(n+1)}}{ \delta G_{ab}} 
\nonumber \\
&=& -\frac{2}{2n} \times 2n \left(-\frac{\lambda}{6} \right)^n \frac{\delta_{ab}}{ N} 
\left( \prod_{j=1}^n T \sum_{\omega_j} \int \! \frac{d^d k_j}{(2\pi)^d} \right) G(\omega_1, {\bf k_1}) 
\nonumber \\
&& \times G(\omega+ \omega_1 -\omega_2, {\bf q+k_1-k_2}) G(\omega_2, {\bf k_2})  \cdot \cdot \cdot 
 G(\omega+ \omega_1 - \omega_n, {\bf q+k_1-k_n}) G(\omega_n,{\bf k_n})
\nonumber \\
&=& -2 \left(-\frac{\lambda}{6} \right)^n  \frac{\delta _{ab}}{ N} \int \! \frac{d^{d+1}k_1}{(2\pi)^{d+1}} \rho(k_1)
\nonumber \\
&& \times \left( \prod_{j=2}^n \int \! \frac{d^{d+1}k_j}{(2\pi)^{d+1}} \int \! \frac{d^{d+1}p_j}{(2\pi)^{d+1}}
(2\pi)^d \delta^{(d)} ({\bf p_j +k_j - q-k_1}) \rho(p_j) \rho(k_j) \right) \times S_{\omega}
\label{eq:pi}
\end{eqnarray} 
with changing variables $\omega = i 2\pi l T \rightarrow q^0$.
As a result we find ${\cal S}'=0$ for its principle values due to the cancellation order by order.  
When singularities of skeletons for $n\geq 3$ remain,
 memory correction terms appear in thermal equilibrium in the final expression
of entropy density $\cal S$ as Refs.~\cite{IKV4,CP1975}.

\end{appendix}
\end{widetext}

\end{document}